\documentclass[preprint]{vgtc}               

\ifpdf
  \pdfoutput=1\relax                   
  \usepackage{graphicx}                
  \DeclareGraphicsExtensions{.pdf,.png,.jpg,.jpeg} 
\else
  \usepackage{graphicx}                
  \DeclareGraphicsExtensions{.eps}     
\fi%

\graphicspath{{figures/}{pictures/}{images/}{./}} 

\usepackage{microtype}                 
\PassOptionsToPackage{warn}{textcomp}  
\usepackage{textcomp}                  
\usepackage{mathptmx}                  
\usepackage{times}                     
\usepackage{cite}                      
\usepackage{tabu}                      
\usepackage{booktabs}                  

\onlineid{1041}

\vgtccategory{Research}
\vgtcpapertype{theory/model}

\vgtcinsertpkg

\title{GlassViz: Visualizing Automatically-Extracted Entry Points for Exploring Scientific Corpora in Problem-Driven Visualization Research}

\author{Alejandro Benito-Santos\thanks{e-mail: abenito@usal.es} %
\and Roberto Ther\'{o}n\thanks{e-mail: theron@usal.es}}
\affiliation{\scriptsize VisUSAL Research Group. Universidad de Salamanca, Spain}

\teaser{
  \centering
  \includegraphics[width=\linewidth]{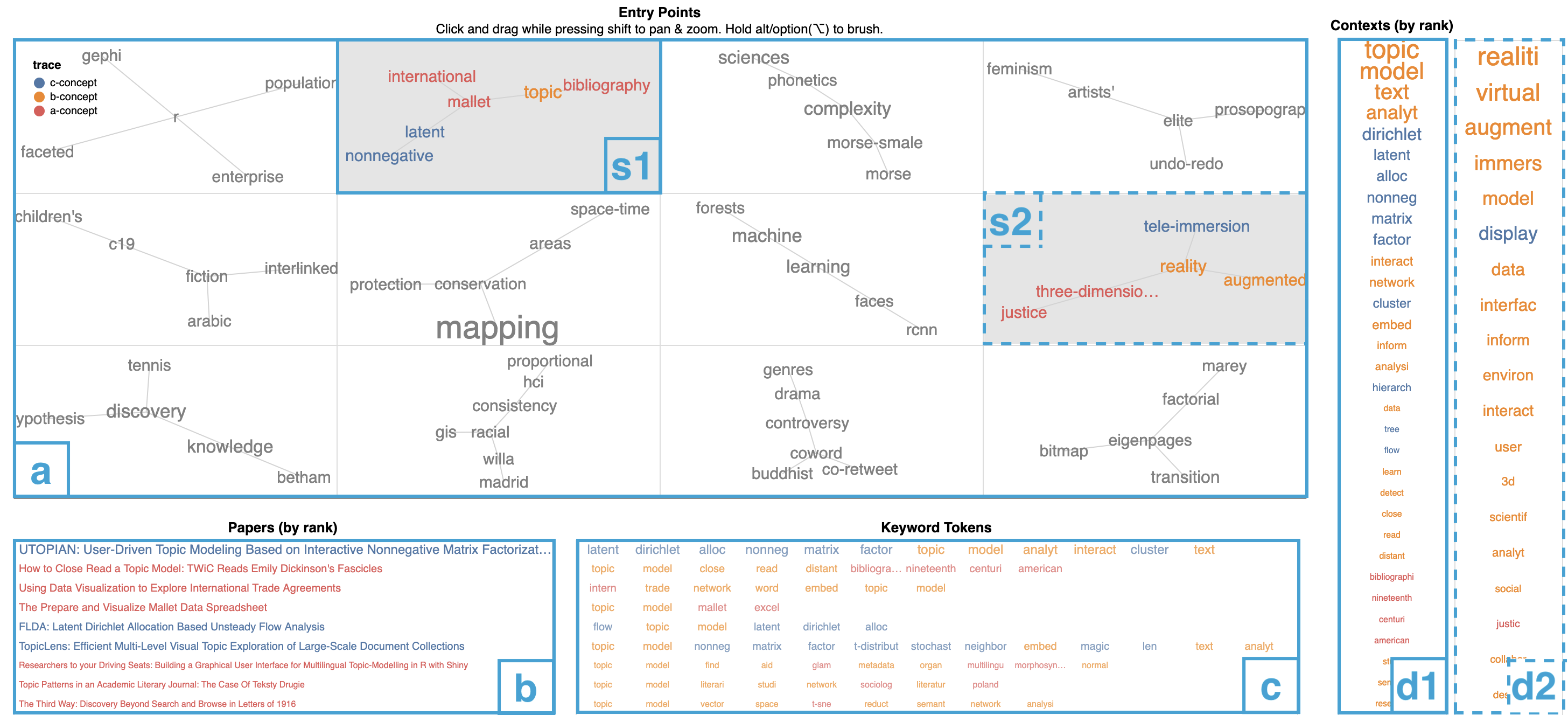}
    \caption{GlassViz interface showing entry points to a corpus of visualization research papers along with related documents and keywords: (a) Quality neighborhoods representing entry points as connected keyword groups. (b) List of documents (showing only the first nine) sorted by number of keyword tokens matching those in selection s.1. (c) Keyword tokens for each document in view b. (d.1) Ranked list of tokens appearing in view c informing of the composition of topics in the entry point selected in s.1. (d.2) State of view d.1 when the selection in view "a" is changed to s.2.}
  \label{fig:glassviz}
}
\abstract{In this paper, we report the development of a model and a proof-of-concept visual text analytics (VTA) tool to enhance document discovery in a problem-driven visualization research (PDVR) context. The proposed model captures the cognitive model followed by domain and visualization experts by analyzing the interdisciplinary communication channel as represented by keywords found in two disjoint collections of research papers. High distributional inter-collection similarities are employed to build informative keyword associations that serve as entry points to drive the exploration of a large document corpus. Our approach is demonstrated in the context of research on visualization for the digital humanities. 
} 

\begin{document}



\maketitle

\section{Introduction} 
\label{sec:intro}
Problem-driven visualization research (PDVR) \cite{simon_bridging_2015} requires intensive collaboration between visualization and domain experts to solve problems in a specific academic discipline such as biology, sports science, computer security, or the humanities. Motivated by the increasing specialization and difficulty of said problems, this collaboration usually materializes in the celebration of workshops, parallel events and micro-conferences (e.g., BioVis, Vis4DH, CityVis, VizSec) and related specialized publication datasets. Setting aside each domain's particularities, these communities generally have to deal with the same typical problems of visualization practice (e.g., dimensionality reduction, hierarchy visualization, or color perception). To obtain insight on these topics and generate novel research ideas\cite{guo_topic-based_2018}, researchers perform literature reviews on other larger datasets of visualization publications in search of techniques conceived in other domains that may assist them in solving specific problems of their own domains. This transference of knowledge between communities of practice is known in human-computer interaction (HCI) and visualization as "methodology transfer" (MT), that is, \textit{“the action of utilizing available models that provide solutions to existing and unsolved problems”} \cite{burkhard_learning_2004, miller_framing_2019}. For example, under this paradigm, a digital humanist focusing on the analysis of digital editions may find interesting a visual algorithm conceived for the analysis of genetic data or vice-versa (e.g., an Arc Diagram \cite{wattenberg_arc_2002-2}). However, the arrival at this kind of findings is seldom straightforward. A first hurdle is related to the lack of linguistic competences \cite{simon_bridging_2015} to formulate queries that serve as entry points \cite{guo_topic-based_2018} to the dataset. To illustrate this situation, take the example of the same digital humanist willing to explore a large corpus of visualization research papers. From previous experience, she knows that the analysis of digital editions is typically related to the concepts of "network analysis" and "graph theory", which are her \textit{entry points} to the dataset. However, she might not be familiar yet with other more specific techniques that could be useful in this context, such as "graph clique" or "persistent homology." Conversely, the authors of papers containing these specific terms might not have chosen to include the more general terms "network analysis" or "graph theory" in their keyword selections for being too obvious and thus uninteresting for the audience addressed initially in their works. Therefore, these publications are effectively invisible to the digital humanist's eyes because she has not yet acquired the necessary vocabulary to formulate an adequate query for this dataset. Irremediably, in a typical setup she will have to start the search by typing keyword(s) she is familiar with, initiating an iterative sensemaking process \cite{pirolli_2005, klein_making_2006}, that will be followed by a faceted browsing of the dataset according to its different dimensions (e.g., authors, keywords, or citations). The situation depicted in this example presents further problems: firstly, searching by general terms will return large document lists with varying degrees of relevance that the researcher needs to inspect and filter individually. Second, the subsequent browsing is performed by manual means following a chain of first-order co-occurrences of metadata items, which may rapidly become a frustrating experience for the user, especially when the data volumes are large. To overcome these issues, we propose a distributional model and a related proof-of-concept (POC) tool that aim to capture similarities between keywords in different domains and to automate the generation of meaningful entry points to a corpus of research papers that needs to be explored. The model and tool are demonstrated in the context of a researcher working at the intersection of visualization and the humanities.

\section{Related Work} 
\noindent
\textbf{Problem-Driven Visualization Research (PDVR):} PDVR brings together domain and visualization experts that collaborate to solve specific, \textit{inherently complex} domain problems. Beyond technical expertise in both domains, some authors have stressed the importance of language to success in interdisciplinary research \cite{bracken_what_2006}. In this regard, Simon et al. explain collaborations in PDVR with a communication model\cite{simon_bridging_2015} in which domain experts generate the \textit{problem space} by providing \textit{data} and \textit{driving problems}, and visualization experts contribute \textit{exploratory data analysis and visualization} techniques defining the \textit{design space}. Following this reasoning, solutions are mappings between the problem and design spaces, and their number is defined by the breadth (or richness) of the communication channel shared by the two teams. More recently, Miller et al.\cite{miller_framing_2019} developed these concepts further in their Methodology Transfer Model (MTM). The MTM incorporates the notions of \textit{similarity} and \textit{alignment} to identify potential MTs between different knowledge domains. Our work employs distributional similarity to extend these theoretical models with other concepts drawn from information science (see next paragraph). \textbf{Literature-Based Discovery (LBD):} LBD is a knowledge extraction technique that \textit{"generates discoveries, or hypotheses, by combining what is already known in the literature."}\cite{thilakaratne_systematic_2019} The concept was introduced in the 1980s by Don R. Swanson, an information scientist known for coining the first form of LBD, the \textit{ABC model} \cite{swanson_fish_1986}. The \textit{ABC model} employs \textit{transitive inference} to unveil non-trivial implicit associations between two disjoint bodies of scientific literature (source and target). It utilizes a simple yet powerful syllogism to pair knowledge fragments: If a term/concept (a-concept) is related to the intermediate term/concept (b-concept) which appears in both the source and target literatures, and the b-concept is related to a c-concept which only appears in the target literature, then we can find a relation, characterized by the b-concept, between the a-concept (which the user is familiar with) and the c-concept (which is new to user). Specifically, we look upon recent work by Thilakaratne et al.\cite{thilakaratne_automatic_2018}, who employ word embeddings to find interesting cross-disciplinary affinities in online paper databases. As opposed to the authors, who employ paper abstracts to generate neural embeddings using the \textit{word2vec} model \cite{mikolov_distributed_2013}, our work relies on author-assigned keywords (hereinafter simply "keywords"), which are descriptive words assigned by the authors to their research papers and have been successfully employed in the past by other researchers to "facilitate the process of understanding differences and commonalities of the various research sub-fields in visualization."\cite{isenberg_visualization_2017}. Also, and despite recent efforts \cite{chuang_without_2012}, the process by which humans extract keywords from academic texts remains mostly unknown \cite{lahiri_replication_2019}. Therefore, keywords model a unique and highly expressive language that serves as the starting point for our study. \textbf{Visual Text Analytics (VTA) of Scientific Literature:} In recent times, some authors have started to incorporate linguistic and sensemaking models into their VTA tools to replicate the typical tasks and goals of exploring scientific texts\cite{federico_survey_2017}. For example, the Action Science Explorer \cite{dunne_rapid_2012} and PaperPoles \cite{he_paperpoles_2019} mimic the sensemaking process of traditional literature reviews. Concretely, PaperPoles supports the browsing of publications in a context-aware environment by requesting positive or negative queries from the user as the application workflow progresses. PaperQuest \cite{ponsard_paperquest_2016} employs a relevance algorithm to rank papers according to the sensemaking process of literature reviews. PaperQuest assumes that the user has one or more \textit{seed papers} at her disposal to start the exploration, a concept that we implemented in GlassViz. Guo et al. \cite{guo_topic-based_2018} propose a two-stage sensemaking framework to discover novel research ideas based on previous work by Pirolli and Card \cite{pirolli_2005}. Wang et al. implement two different logic flows in their system (author-based and citation-based) to mirror the traditional literature review process \cite{wang_guided_2016}. To the best of our knowledge, GlassViz is the first VTA tool to incorporate the sensemaking model followed by interdisciplinary visualization researchers using an LBD workflow.

\section{Data Processing}
\label{sec:data}
We selected two research paper collections as the S and T literatures in our LBD setup. T-Literature (VIS4DH dataset), representing the target domain that solutions need to be imported to, comprises 221 papers on visualization for the Digital Humanities (DH)\cite{benito-santos_data-driven_2020}. S-Literature (VIS dataset) is a set of 2117 visualization publications that have appeared at the IEEE Visualization set of conferences: InfoVis, SciVis, VAST and Vis between the years 1991-2018 \cite{isenberg_vispubdata.org_2017}. Keywords were extracted from each document, tokenized and translated into their American English forms when necessary. Tokens matching NLTK's list of English stop words (e.g., "and" or "of") were removed from further analysis, which yielded a total of 3403 different tokens. Next, each token was light-stemmed using the Porter algorithm. Given that keywords are a very sparse feature of scientific papers, the stemming procedure had the positive effect of compressing the input vocabulary (from 3403 to 2720 tokens) by linking redundant forms together under the same root (e.g., "filtering," "filters" and "filtered" under "filter"). In addition, and despite certain limitations that we discuss in \autoref{sec:conclusion}, the stemming algorithm helped relate documents referring to the same high-level concepts requiring minimal human intervention (e.g., a manual classification\cite{isenberg_vispubdata.org_2017}). Finally, we removed uninteresting tokens with inverse document frequency (IDF) of less than 1.0, resulting in only one token ("visual") being discarded. Each document was treated as a bag-of-(key)word tokens defining a vocabulary composed of three disjoint sets as per Swanson's ABC model: $V_{a}$ (a-concepts, or tokens appearing \textit{exclusively} in the VIS4DH dataset), $V_{c}$ (c-concepts, or tokens appearing \textit{exclusively} in the VIS dataset), and $V_{b}$ (b-concepts, or tokens appearing in both datasets). In the end, the vocabulary sizes obtained were: $|V_{a}| = 259$, $|V_{b}| = 302$, and $|V_{c}| = 2159$.

\section{System Design}
\subsection{Tasks and Design Goals}
Our approach relies on the extraction of entry points to guide the exploration of a scientific corpus. The extraction of the entry points is based on the following assumptions: at the beginning of this study, we observed that researchers participating in PDVR internally follow an MTM that is mainly driven by their previous experience in other projects and domains. Here, the expert initially analyzes the problem and breaks into its constituent parts, leading to a set of themes that are matched against previous grounded knowledge. In this mental process, candidate solutions are detached from the original problem's domain and matched against the new domain in search of viable solutions. The most \textit{similar} solutions are then implemented to obtain preliminary insight into the data, which is often necessary to promote discussions between stakeholders and advance the project at its early stages. Later in the design process, the team may decide to modify and/or combine these initial solutions to provide a visualization that aligns better with the data and tasks of the problem at hand\cite{miller_framing_2019}. Motivated by the presented situation, we extracted the following design goals and related questions at the beginning of the study, which ultimately drove the design of our distributional model and POC tool: \textbf{DG.1}: Motivate a personalized exploration of scientific corpora that is tailored to the user’s research aims. \textit{"What kind of knowledge does the user want to extract from a dataset?", "What can a user learn from the dataset that is useful for solving a particular domain problem?"} \textbf{DG.2}: Potentiate the discovery of methodologies that could potentially be transferred from other existing design spaces to the source domain. \textit{"How can we measure the degree of transferability of solutions conceived in other knowledge domains?"} \textbf{DG.3}: Accelerate sensemaking and language acquisition in the context of PDVR. \textit{"What are the most informative terms that best describe a dataset according to the user’s level of expertise and grounded knowledge?", What themes are especially interesting for the user?", "How can they be presented in the best possible manner to augment their comprehensibility?"} \textbf{DG.4}: Provide a reading order for discovered documents. \textit{What documents are the most important in the collection for the user?}

\subsection{Theoretical Model}
\label{sec:model}
\begin{figure}
  \centering
  \includegraphics[width=.95\linewidth]{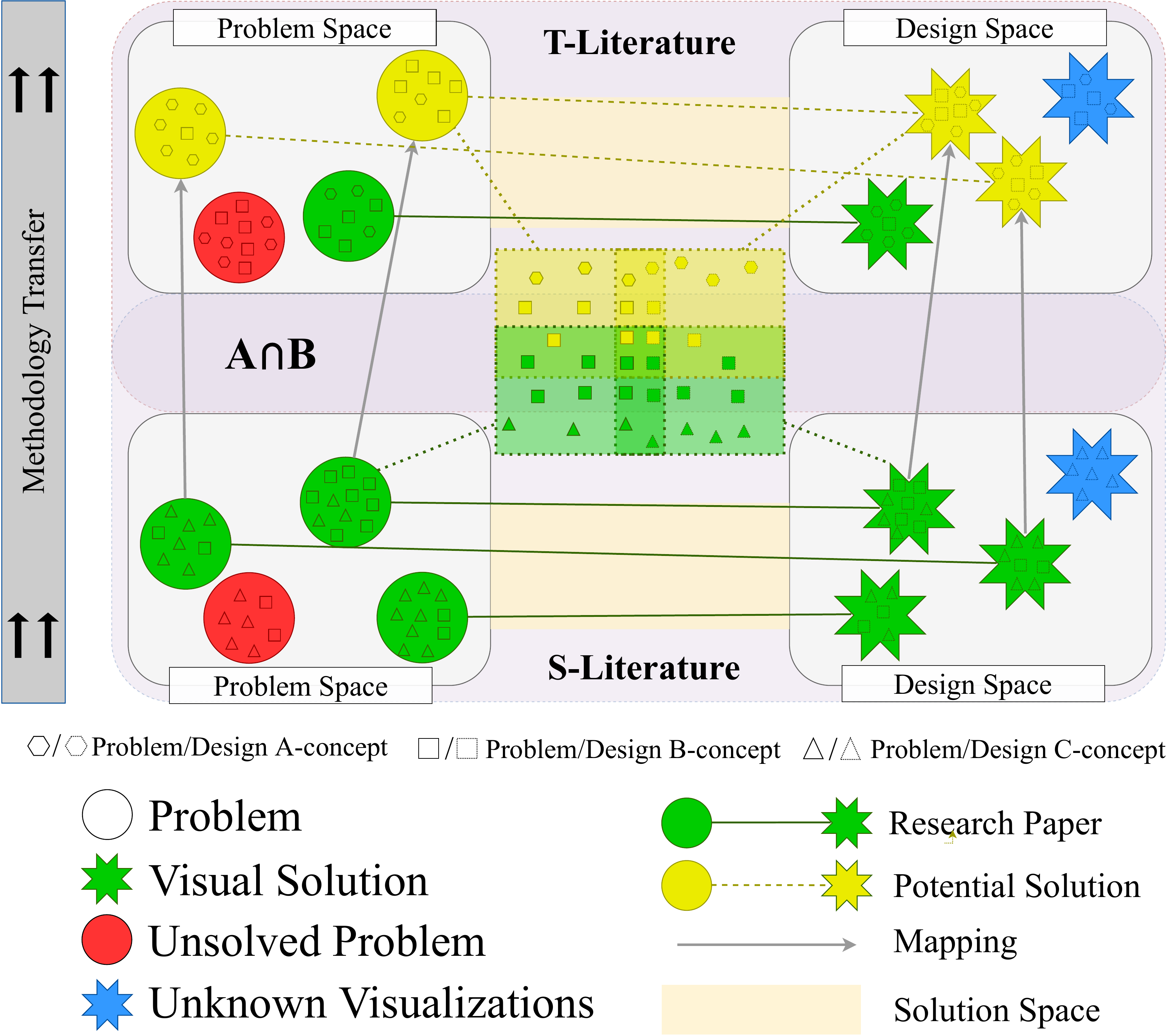}
  \caption{\label{fig:methodology-transfer-abc}
           Methodology Transfer Model (MTM) adapted from Miller et al. \cite{miller_framing_2019}. The model maps problems and designs found in two disjoint bodies of literature and it is augmented with concepts drawn from Swanson's ABC Model for Literature-Based Discovery to automate the discovery of candidate MTs and to provide the user with informative entry points to the S-Literature.}
\end{figure}

Our theoretical model (\autoref{fig:methodology-transfer-abc}) combines Swanson’s and Miller et al.’s models to build automatic entry points that resemble the researchers’ sensemaking model and assist them in the task of mapping problem and design spaces in different domains and bodies of literature. 
According to Simon et al.\cite{simon_bridging_2015}, the problem space is defined by application domains and their data, whereas the design space comprises analytical tasks and visualizations. In the diagram, we depict the idea that valid MTs consist of a series of concepts specific to each domain (a- and c-concepts) and a variable number of techniques that address a generic, high-level problem in the visualization domain (b-concepts). Thus, as per Swanson's model, solutions (or papers) in the T-literature link problems and designs containing only a- and b-concepts, while those in the S-literature contain only c- and b-concepts. Then, it should be theoretically possible to deduct recurrent terms of potential solutions by analyzing the distribution of terms in existing solutions documented in the literature in other domains and relating them to the problem(s) at hand using high-order co-occurrence. This idea is depicted in the Venn diagram at the center of the image. At the intersection of the four sets, the core terms of the elements in the four spaces meet, giving clues about the descriptions of potential solutions. Besides, more potential solutions could be found by following chains of co-occurrence that led to peripheral intersection spaces. As we explain in the next section, our proposed model captures high-order co-occurrence of concepts for enhancing the document exploration process. 
\subsection{Keyword Embeddings}
We rely on the generation of keyword embeddings for detecting distributional similarities between problems, data, tasks, or visualizations in the S- and T-literatures. These embeddings were generated by following the method proposed by Levy et al.\cite{levy_neural_2014}, which requires minimal hyper-parameter tuning and they are known to excel at word similarity tasks \cite{levy_neural_2014, levy_improving_2015}. Initially, the method relies on an initial pointwise mutual information (PMI) matrix that encodes the probability for a pair of keyword tokens to be seen together in a document with respect to the probability of seeing those two same tokens in the union of the two corpora (see \autoref{eq:pmi}). For all keyword token pairs in the S and T literatures, each cell $M_{i,j}$ represents the log odds ratio of $w_{i}$ (a keyword) and $c_{j}$ (any other keyword appearing with \textit{w} in a document D, its context) joint probability and the product of their marginal probabilities. The marginal probabilities were empirically obtained from the corpora by counting the number of occurrences of each token divided by the union size of the document collections.
\begin{equation}
\label{eq:pmi}
    PMI(w,c) = \log\frac{P(w,c)}{P(w)P(c)} = \log\frac{\#(w,c) \cdot |D|}{\#(w) \cdot \#(c)}
\end{equation}
Given that PMI can be $-inf$ for pairs of tokens that were never seen in the corpus, it is customary to use the positive version of the PMI matrix that is defined as:
\begin{equation}
\label{eq:ppmi}
    PPMI(w,c) = max(PMI(w,c),0)
\end{equation}
Following recommendations in the literature\cite{levy_improving_2015, benito-santos_cross-domain_2019}, we applied a light smoothing with $\alpha = 0.95$(\autoref{eq:alpha}) to counterbalance the PMI bias towards infrequent events (note that the alpha factor is a corpus-dependent parameter and was manually adjusted).
\begin{equation}
\label{eq:sppmi}
    SPPMI(w,c) = \log\frac{\hat{P}(w,c)}{\hat{P}(w)\hat{P}_{\alpha}(c)} \\
\end{equation}
where the smoothed unigram distribution of the context is:
\begin{equation}
\hat{P}_{\alpha}(c) = \frac{\#(c)^{\alpha}}{\sum_{c}{\#(c)^{\alpha}}}
\label{eq:alpha}
\end{equation}
To capture high-order co-occurrence and to generate dense keyword vectors from the sparse SPPMI matrix was factorized into the product of three matrices by applying a non-parametric algebraic method, SVD, which was popularized in the NLP community with Latent Semantic Analysis (LSA)\cite{deerwester_indexing_1990, levy_neural_2014}. If the SPPMI matrix is the matrix $M$, SVD decomposes $M$ into the product of three matrices $U \Sigma V^{T}$, where $U$ and $V$ are orthonormal ($U^{T} U = V^{T} V = I$) and $\Sigma$ is a diagonal matrix of sorted singular values of the same rank \textit{r} as the input matrix. Then, our resulting vector space model (VSM) is formed by dense keyword embeddings resulting from keeping only the first $k$ columns in $U$ ($k=50$ in our case).

\section{GlassViz}
In this section, we describe the design decisions that drove the development of our prototype tool by carrying an experiment using the datasets introduced in \autoref{sec:data}. Our approach is centered around the qualitative inspection of \textit{quality} local neighborhoods of a-concepts that were derived using a \textit{cosine metric}\cite{thilakaratne_automatic_2018}. According to the literature, it is customary to select between 3 and 5 nearest neighbors for this task (see \cite{heimerl_interactive_2018}, section 4.1.1) Thus, we decided to extract the 4 nearest neighbors for each a-concept $t_{a}$ in $V_{a}$. Tokens represented by very similar vectors ($sim(t_{a}, t_{b}) \leq 0.01$) and thus displaying identical nearest neighbors were considered redundant for the purpose of this task and thus were removed (488 in total). Quality neighborhoods were defined as those containing significant similarities between a- and c- concepts and were identified by two criteria: (1) the neighborhood included at least one c-concept, and (2) when criterion 1 was met, the similarity between the a-concept and its nearest c-concept in the neighborhood fell within the first quartile of all highest similarities ($dist(t_{a}, t_{c}) <= Q_{1}$, with $Q_{1} = 0.2451$), which yielded 15 quality neighborhoods. To relate neighborhoods representing similar themes, neighborhoods with common terms were merged, resulting in 12 distinct entry points. Finally, we wanted to display the neighborhood's embedding subspaces defined by each entry point in the best possible manner to motivate a gradual transition from familiar a-concepts to interesting, possibly unknown c-concepts. This implied representing not only similarities between the nearest neighbors and the a-concept originating the neighborhood but also showing similarities among neighbors. To this end, we relied on a graph scaling technique, pathfinder networks (PFNETs)\cite{schvaneveldt_pathfinder_1990} that was applied to the complete similarity subgraphs formed by terms on each entry-point. PFNETs are well-known in the visualization and information theory literature \cite{chen_visualising_1999, chen_visualizing_2001, benito-santos_cross-domain_2019} for their suitability to capture underlying knowledge structures (\textbf{DG.1}) and for motivating a fast vocabulary learning (\textbf{DG.3}) with a minimal cognitive gap. This is achieved by pruning graph edges that are not on shortest paths according to two parameters $q$ (the number of indirect proximities considered to build the PFNET) and $r$ (the metric used to compute pairwise similarities) \cite{schvaneveldt_pathfinder_1990, chen_visualising_1999, chen_visualizing_2001}. Concretely, we calculated minimum spanning trees (MSTs), the most concise form of a PFNET ($q=n-1, r=\infty$), for the 12 complete subgraphs. Following recommendations in the literature \cite{chen_visualising_1999}, each PFNET was plotted using a force-directed algorithm \cite{fruchterman_graph_1991-2} that placed nodes displaying high pairwise cosine similarities closer in the chart. In this representation, the nodes depict a-,b-, or c-concepts as per Swanson's ABC model. Each MST portrays an exploration path (or entry point) to the VIS dataset that can be inspected individually in the designated areas of view 1.a (see \autoref{fig:glassviz}). A total of 29 a-concepts (red), 16 b-concepts (yellow) and 19 c-concepts (blue) were captured. Each node shows a text label containing the most common form of the corresponding token and whose size encodes the token(s)'s absolute frequency in the union of the two literatures. In view 1.b, documents in the current selection are listed in descending order of number of keyword tokens matching those in the current selection (\textbf{DG.4}). The number of documents containing any of the terms captured by the entry points was 69 for the T-Literature A and 297 for the S-Literature (31.22\% and 14.03\% coverage, respectively). To the right of view 1.b., view 1.c shows keyword tokens for each document shown in view 1.b, whereas view 1.d1 (and 1.d2 for selection s2) aggregates and presents these tokens in a rank frequency list. 

By visually inspecting each of the 12 entry points in view 1.a (\textbf{DG.1}), the user can recognize interesting inter-collection distributional similarities between concepts describing application areas, domain problems, analytical tasks/techniques and visualizations as per the model introduced in \autoref{sec:model}. The entry points can be further inspected using a brushing+linking interaction technique For example, when brushing the entry point in selection $s1$ of \autoref{fig:glassviz}, views 1.b, 1.c and 1.d1 are updated. Here, the entry point introduces two c-concepts ("nonnegative" and "latent") that are related by their distributional similarity to two DH-specific problems ("bibliography" and "international) and a technique ("mallet") depicted by the a-concepts in red in the diagram. To get a better understanding of the entry point's underlying theme and concepts (\textbf{DG.3}), the user could look at view 1.d1 to discover the most frequent tokens ("topic," "model," "text," "analyt," "dirichlet," etc.) found in documents matching any of the entry point's concepts shown in s1, allowing a first rapid interpretation of the theme. By interacting with the items in view 1.b, the user could retrieve in a pop-up view multiple related metadata to a document; i.e., title, authors list, publication year/venue and number of matching keywords with the entry point. In the same view, it can be observed that the three a-concepts in s1 can be traced to three documents in the VIS4DH dataset describing two domain problems (the analysis of international trade agreements and bibliographic works, respectively) and a domain-specific analysis tool, a wrangling Excel script for a popular NLP toolkit among DH practitioners. Similarly, the two c-concepts "nonnegative" and "latent" can be traced to three other documents in the VIS dataset and reconstructed by the user to "nonnegative matrix factorization" and "latent dirichlet analysis," introducing potentially interesting analysis techniques (\textbf{DG.2}). The same workflow could be applied to any other entry point of view 1.a, for example to the one depicted in selection s2. This entry point relates the domain problem of "social justice" to the a-concept "tele-immersion" under the background theme of virtual and augmented reality (view 1.d2). 

\section{Conclusions, Limitations and Future Work}
\label{sec:conclusion}
We have presented a model and a related VTA prototype aimed at accelerating the process of knowledge and language acquisition in PDVR. By modeling the distribution of keywords defining the interdisciplinary communication channel as documented by research papers found in two disjoint bodies of literature, we were able to generate entry points that motivated a personal exploration of a corpus of visualization papers according to the researcher's particular needs and expectations and that required minimal user intervention. However, we identified the existence of certain limitations in our approach that are discussed hereafter: firstly, the stemming algorithm employed to compress the input data produced some false positives that are difficult to avoid by automatic means. Concretely, this side effect could be observed in cases where keywords with different meanings were reduced to the same lexical form, for example in the tokens \textit{factory} (from "smart factory") and \textit{factorial} (from "factorial analysis"). Also, \textit{GlassViz} does not allow the interactive tuning of certain parameters such as the number $k$ of singular values, the smoothing alpha factor $\alpha$, or the similarity thresholds set to detect redundant vectors and quality neighbors. To resolve these and other issues, we plan to incorporate direct manipulation techniques \cite{el-assady_semantic_2020} in the future. Furthermore, the tokenization of keywords increased the difficulty in interpreting the entry points' background themes, a limitation that could be resolved by employing auxiliary n-gram statistics\cite{chuang_termite:_2012} to assist the user in reconstructing the original phrases. 

\acknowledgments{
The authors want to thank the three anonymous reviewers for their helpful comments. This work was supported by
a grant from the Spanish Ministry of Economic Affairs and Digital Transformation under the EU CHIST-ERA agreement (PCIN-2017-064).}

\bibliographystyle{abbrv-doi}

\bibliography{template}
\end{document}